\documentclass[amssymb,twocolumn,eqsecnum,aps]{revtex4}
\usepackage{amsmath}
\usepackage{enumerate}
\usepackage{graphicx}
\begin{document}
\def \brho {{\rho} \hskip -4.5pt { \rho}}
\def \bnab {{\bf\nabla}\hskip-8.8pt{\bf\nabla}\hskip-9.1pt{\bf\nabla}}
\title{The Role of Conservation Principles in the Abraham--Minkowski Controversy}
\author{Michael E. Crenshaw}
\email{michael.e.crenshaw4.civ@mail.mil}
\affiliation{US Army RDECOM, Aviation and Missile Research, Development, and Engineering Center, Redstone Arsenal, AL 35898, USA}
\begin{abstract}
The Abraham--Minkowski controversy refers to a long-standing inability
to adequately address certain issues involving the conservation of
the momentum of an electromagnetic field in a linear dielectric medium.
We apply the usual assumption of a material subsystem that couples to
the electromagnetic subsystem such that the total energy and total
momentum are conserved.
We then construct the total energy--momentum tensor from the total
energy density and the total momentum density.
Applying conservation principles to the total energy--momentum tensor,
we construct the tensor energy--momentum continuity equation.
We show that one of the components of the tensor energy--momentum
continuity equation, the energy continuity equation, is manifestly
false.
We conclude that the Abraham--Minkowski controversy is unresolvable
because the extant principles of conservation are inconsistent in a
simple linear dielectric medium.
\hfill
\end{abstract}
\date{\today}
\maketitle
\par
\section{Introduction}
\par
The century-long history of the Abraham--Minkowski controversy
\cite{BIMin,BIAbr,BIPfei,BIAMC2,BIAMC3,BIAMC4,BIAMC5,BIKemplatest,BIObuk,BIMuka,BIBarnL,BIGord,BIBahder,BIJMP}
is a search for some provable description of momentum and momentum 
conservation for electromagnetic fields in dielectric media.
A wide variety of physical principles have been applied to establish the
priority of one type of momentum over another, or to establish that the
Abraham and Minkowski formulations are equally valid.
Typically, one assumes some fundamental physical principle or law and
the correctness of the results are affirmed by the fundamental nature of
the laws that are used as the basis of the analyses, such as the
macroscopic Maxwell equations,
the Lorentz dipole force,
symmetrized Minkowski tensor,
the constancy of the center-of-mass energy velocity,
Lorentz invariance,
or spatially averaged microscopic fields.
In order to satisfy golobal conservation principles, an appropriate material
momentum must accompany these electromagnetic momentums such that the
sum of the electromagnetic and material momentums is the total momentum
that is conserved in a closed system.
\par
For a thermodynamically closed system consisting of a propagating light
field and an antireflection-coated simple linear dielectric, the total
momentum \cite{BIPfei,BIGord,BIBahder,BIJMP},
\begin{equation}
{\bf G}_T=\int_{\sigma} {\bf g}_T dv
=\int_{\sigma}\frac{n{\bf E}\times{\bf B}}{c} dv \, ,
\label{EQf1.01}
\end{equation}
is temporally invariant when the total momentum density ${\bf g}_T$ is
integrated over all-space $\sigma$.
Global conservation of the total momentum ${\bf G}_T$,
Eq.~(\ref{EQf1.01}), has been documented for an antireflection-coated
block of a simple linear dielectric material situated in free-space that
is illuminated by a quasimonochromatic field \cite{BIBahder,BIJMP}.
The modern resolution of the Abraham--Minkowski momentum controversy is
to adopt a scientific conformity in which the Minkowski momentum,
\begin{equation}
{\bf G}_M=\int_{\sigma} {\bf g}_M dv
=\int_{\sigma}\frac{{\bf D}\times{\bf B}}{c} dv \, ,
\label{EQf1.02}
\end{equation}
and the Abraham momentum,
\begin{equation}
{\bf G}_A=\int_{\sigma} {\bf g}_A dv
=\int_{\sigma}\frac{{\bf E}\times{\bf H}}{c} dv \, ,
\label{EQf1.03}
\end{equation}
are both ``correct'' forms of electromagnetic momentum with the
understanding that neither is the total
momentum \cite{BIPfei,BIAMC2,BIAMC5,BIBarnL}.
Then, either the Minkowski momentum or the Abraham momentum can be
adopted as the momentum of the electromagnetic field as long as that
momentum is accompanied by the appropriate material momentum,
${\bf G}_T-{\bf G}_A$ or ${\bf G}_T-{\bf G}_M$.
In the most general interpretation of the consensus resolution of the
Abraham--Minkowski controversy, only the total momentum has physical
meaning and the total momentum can be divided into arbitrary field
and material components \cite{BIPfei}.
\par
It is well known that a material momentum accompanies the
electromagnetic momentum in a dielectric and that their sum, the total
momentum, is conserved.
The components of the total momentum density, along with the total
energy density, are elements of the total energy--momentum tensor.
In this article we report that the tensor total energy--momentum
continuity equation that is constructed from the total energy--momentum
tensor is false.
The Abraham--Minkowski controversy cannot be resolved using the existing
theory because the principles of conservation are inconsistent in a
linear dielectric medium.
\par
\section{Energy--momentum tensors}
\par
The theory of continuum electrodynamics is based on the macroscopic
Maxwell equations
\begin{subequations}
\begin{equation}
\nabla\times{\bf H}-\frac{1}{c}\frac{\partial{\bf D}}{\partial t}=0
\label{EQf2.01a}
\end{equation}
\begin{equation}
\nabla\cdot{\bf B}=0
\label{EQf2.01b}
\end{equation}
\begin{equation}
\nabla\times{\bf E}+\frac{1}{c}\frac{\partial{\bf B}}{\partial t}=0
\label{EQf2.01c}
\end{equation}
\begin{equation}
\nabla\cdot{\bf D}=0 
\label{EQf2.01d}
\end{equation}
\label{EQf2.01}
\end{subequations}
and the constitutive relations
\begin{subequations}
\begin{equation}
{\bf D}= \varepsilon {\bf E}
\label{EQf2.02a}
\end{equation}
\begin{equation}
{\bf B}= \mu {\bf H}
\label{EQf2.02b}
\end{equation}
\label{EQf2.02}
\end{subequations}
for an absorptionless linear medium.
For clarity and concision, we treat a limiting case in which the pulse
is sufficiently monochromatic and the center frequency of the exciting
field is sufficiently far from material resonances that dispersion can
be treated parametrically and otherwise ignored.
\par
The energy and momentum continuity equations are the counterparts of
conservation laws when the system consists of a continuous flow 
rather than localized and enumerated discrete particles.
In the particular case of macroscopic electromagnetic energy and
momentum, the continuity equations are easily derived
from the macroscopic Maxwell equations,
Eqs.~(\ref{EQf2.01})--(\ref{EQf2.02}).
We subtract the scalar product of Eq.~(\ref{EQf2.01a}) with ${\bf E}$
from the scalar product of Eq.~(\ref{EQf2.01c}) with ${\bf H}$ and apply
a common vector identity to produce
\begin{equation}
\frac{1}{c}\left (
{\bf E}\cdot\frac{\partial{\bf D}}{\partial t}
+{\bf H}\cdot \frac{\partial{\bf B}}{\partial t}
\right )
+\nabla\cdot ( {\bf E}\times{\bf H})=0 \, .
\label{EQf2.03}
\end{equation}
We define the macroscopic electromagnetic energy density
\begin{equation}
\rho_e= \frac{1}{2}\left ( \varepsilon {\bf E}^2+\mu {\bf H}^2 \right )
\label{EQf2.04}
\end{equation}
and the Poynting energy flux vector
\begin{equation}
{\bf S}=c {\bf E}\times{\bf H}\, ,
\label{EQf2.05}
\end{equation}
as usual, to obtain Poynting's theorem
\begin{equation}
\frac{\partial \rho_e}{\partial t}+ \nabla\cdot {\bf S} = 0
\label{EQf2.06}
\end{equation}
for continuity of electromagnetic energy in a linear medium.
We can also form
\begin{equation}
\frac{\partial}{\partial t}\frac{{\bf D}\times{\bf B}}{c} 
=-{\bf D}\times(\nabla\times{\bf E})-{\bf B}\times(\nabla\times{\bf H})
\label{EQf2.07}
\end{equation}
from the difference of cross products of Eqs.~(\ref{EQf2.01a}) and
(\ref{EQf2.01c}) with macroscopic fields.
In what follows, the index convention for Greek letters is that they
belong to $\{0,1,2,3\}$ and lower case Roman indices from the middle
of the alphabet are in $\{1,2,3\}$.
Defining the stress-tensor
\begin{equation}
W_{ij}=-E_i D_j-H_iB_j+
\frac{1}{2} ({\bf E}\cdot{\bf D}+{\bf H}\cdot{\bf B})\delta_{ij}
\label{EQf2.08}
\end{equation}
and the Minkowski momentum density
\begin{equation}
{\bf g}_M=\frac{{\bf D}\times{\bf B}}{c} 
\label{EQf2.09}
\end{equation}
yields the momentum continuity equation
\begin{equation}
\frac{\partial{\bf g}_M}{\partial t}+\nabla\cdot{\bf W}
=-(\nabla\varepsilon)\frac{{\bf E}^2}{2}
-(\nabla \mu) \frac{{\bf H}^2}{2}\, .
\label{EQf2.10}
\end{equation}
As a matter of linear algebra, we can write the energy continuity
equation, Eq.~(\ref{EQf2.06}),
and the three scalar equations from the momentum continuity equation,
Eq.~(\ref{EQf2.10}), as a single matrix continuity equation
\begin{equation}
\partial_{\beta}T_M^{\alpha\beta}=f^{\alpha}_M \,,
\label{EQf2.11}
\end{equation}
where
\begin{equation}
\partial_{\beta}=
\left ( \frac{\partial}{\partial (ct)},
\frac{\partial}{\partial x},\frac{\partial}{\partial y},
\frac{\partial}{\partial z} \right ) \,
\label{EQf2.12}
\end{equation}
is the four-divergence operator,
\begin{equation}
{\bf f}_M
=-\frac{\nabla \varepsilon}{\varepsilon}\frac{{\bf E}\cdot{\bf D}}{2}
- \frac{\nabla \mu}{\mu} \frac{{\bf H}\cdot{\bf B}}{2}
\label{EQf2.13}
\end{equation}
is the Minkowski force density that is a source, or sink, of 
electromagnetic momentum for the field.
(The force density on the dielectric is the Helmholtz force
density ${\bf f}_H=- {\bf f}_M$.)
Also, $f^{\alpha}_M=(0,{\bf f}_M)$ is the Minkowski four-force density,
and
\begin{equation}
T_M^{\alpha\beta}
\!\! = \!\!
\left [
\begin{matrix}
\frac{1}{2}({\bf D}\cdot{\bf E}+{\bf B}\cdot{\bf H})
\!&({\bf E}\times{\bf H})_1 \!& ({\bf E}\times{\bf H})_2 
\!&({\bf E}\times{\bf H})_3
\cr
({\bf D}\times{\bf B})_1    &W_{11}      &W_{12}      &W_{13}
\cr
({\bf D}\times{\bf B})_2    &W_{21}      &W_{22}      &W_{23}
\cr
({\bf D}\times{\bf B})_3    &W_{31}      &W_{32}      &W_{33}
\cr
\end{matrix}
\right ] 
\label{EQf2.14}
\end{equation}
is, by construction, a four-by-four matrix.
The Minkowski matrix differential continuity equation,
Eq.~(\ref{EQf2.11}), has the outward appearance of being a tensor
energy--momentum continuity equation and it was assumed to be so.
Then, the matrix $T^{\alpha\beta}_M$, Eq.~(\ref{EQf2.14}), became known
as the Minkowski energy--momentum tensor.
\par
The origin story of the Abraham-Minkowski controversy is that the
Minkowski energy--momentum tensor, Eq.~(\ref{EQf2.14}), is not
diagonally symmetric.
Motivated by the need to preserve the principle of conservation of
angular momentum, Abraham symmetrized the energy--momentum tensor
by re-defining the linear momentum density,
\begin{equation}
{\bf g}_A={\bf E}\times{\bf H}/c \, ,
\label{EQf2.15}
\end{equation}
to be proportional to the Poynting vector.
Substituting the Abraham momentum density into Eq.~(\ref{EQf2.10}),
we can construct the matrix continuity equation
\begin{equation}
\partial_{\beta}T_A^{\alpha\beta}=f^{\alpha}_A \,,
\label{EQf2.16}
\end{equation}
where
\begin{equation}
f_A^{\alpha}
=\left (
0, -\frac{\nabla \varepsilon}{\varepsilon}\frac{{\bf E}\cdot{\bf D}}{2}
- \frac{\nabla \mu}{\mu} \frac{{\bf H}\cdot{\bf B}}{2}
+\frac{\partial}{\partial t}
\frac{(1-n^2){\bf E}\times{\bf H}}{c} 
\right )
\label{EQf2.17}
\end{equation}
is the Abraham four-force,
\begin{equation}
T_A^{\alpha\beta} \!\! = \!\!
\left [
\begin{matrix}
\frac{1}{2}({\bf D}\cdot{\bf E}+{\bf B}\cdot{\bf H})
\!&({\bf E}\times{\bf H})_1 \!& ({\bf E}\times{\bf H})_2 
\!&({\bf E}\times{\bf H})_3
\cr
({\bf E}\times{\bf H})_1    &W_{11}      &W_{12}      &W_{13}
\cr
({\bf E}\times{\bf H})_2    &W_{21}      &W_{22}      &W_{23}
\cr
({\bf E}\times{\bf H})_3    &W_{31}      &W_{32}      &W_{33}
\cr
\end{matrix}
\right ] 
\label{EQf2.18}
\end{equation}
is known as the Abraham energy--momentum tensor,
and $\partial_{\beta}$ is again the four-divergence operator.
\par
\section{Conservation}
\par
It has been established in the scientific literature that neither
the Abraham momentum nor the Minkowski momentum is conserved in an
isolated system \cite{BIPfei}.
Although the definitions of the linear momentum and the energy--momentum
tensor have been the most contentious issues, the main challenge of the
Abraham--Minkowski controversy is to construct a tensor energy--momentum
continuity equation.
The conditions \cite{BILL}
\begin{subequations}
\begin{equation}
\partial_{\beta} T^{\alpha\beta} =0
\label{EQf3.01a}
\end{equation}
\begin{equation}
T^{\alpha\beta} = T^{\beta\alpha}
\label{EQf3.01b}
\end{equation}
\begin{equation}
\int_{\sigma} T^{\alpha 0} dv
\;\;{\rm is}\;{\rm temporally}\;{\rm invariant}
\label{EQf3.01c}
\end{equation}
\label{EQf3.01}
\end{subequations}
insure that the energy and momentum conservation laws are satisfied
for an unimpeded continuous flow of electromagnetic
radiation.
Due to the arbitrary size of the dielectric, we can invoke the limit
that the gradient of the material properties $n$, $\varepsilon$, and
$\mu$ are sufficiently small that source terms involving these gradients
can be neglected.
In this limit, Eq.~(\ref{EQf2.11}) becomes
\begin{equation}
\partial_{\beta} T_M^{\alpha\beta} =0
\label{EQf3.02}
\end{equation}
and has the appearance of the tensor energy--momentum continuity 
equation in the form of Eq.~(\ref{EQf3.01a}),
where $T^{\alpha\beta}=T_M^{\alpha\beta}$.
In fact, Eq.~(\ref{EQf3.02}) proves that the Minkowski momentum,
Eq.~(\ref{EQf1.02}), is conserved.
This result is contradicted by Eq.~(\ref{EQf3.01c}) because global
conservation principles prove that the Minkowski momentum is not
conserved for the thermodynamically closed system that is being
treated here \cite{BIPfei,BIGord,BIBahder,BIJMP}.
\par
There exists an accepted procedure to remove the contradiction in which
the energy--momentum tensor of a material subsystem is added to 
the energy--momentum tensor of an electromagnetic subsystem
\cite{BIPfei,BIAMC2,BIAMC3,BIAMC4,BIAMC5,BIKemplatest,BIObuk,BIMuka,BIBarnL,BIGord}.
The total energy--momentum tensor is
\begin{equation}
T^{\alpha\beta}=T_{em}^{\alpha\beta} +T_{matter}^{\alpha\beta} \, .
\label{EQf3.03}
\end{equation}
The electromagnetic subsystem energy--momentum tensor is typically the
Abraham tensor or the Minkowski tensor.
Because the total energy--momentum tensor is unique, the selection of
an electromagnetic tensor determines the material tensor.
By definition, the total energy and the total linear momentum are 
unique conserved quantities that satisfy the condition
Eq.~(\ref{EQf3.01c}).
Based on global conservation principles, we can take the total energy
density
\begin{equation}
T^{00}=T^{00}_{em}+T^{00}_{matter}= \rho_e=
(n^2{\bf E}^2+{\bf B}^2)/2
\label{EQf3.04}
\end{equation}
and the components of the total momentum density
\begin{equation}
T^{i0}=T^{i0}_{em}+T^{i0}_{matter}=
{{\bf g}_T}_i=(n {\bf E}\times{\bf B}/c)_i
\label{EQf3.05}
\end{equation}
as given quantities.
Applying the total energy density, Eq.~(\ref{EQf3.04}),
the total momentum density, Eq.~(\ref{EQf3.05}),
and conservation of angular momentum, Eq.~(\ref{EQf3.01b}), one
constructs the total energy--momentum tensor
$$
T^{\alpha\beta} \! = \!
$$
\begin{equation}
\left [
\begin{matrix}
\frac{1}{2}(n^2{\bf E}^2+{\bf B}^2)
\!&(n{\bf E}\times{\bf B})_1 \!& (n{\bf E}\times{\bf B})_2 
\!&(n{\bf E}\times{\bf B})_3
\cr
(n{\bf E}\times{\bf B})_1    &Y_{11}      &Y_{12}      &Y_{13}
\cr
(n{\bf E}\times{\bf B})_2    &Y_{21}      &Y_{22}      &Y_{23}
\cr
(n{\bf E}\times{\bf B})_3    &Y_{31}      &Y_{32}      &Y_{33}
\cr
\end{matrix}
\right ] \, .
\label{EQf3.06}
\end{equation}
The components of $Y$ are left unspecified as they are not needed for
the present discussion.
Substituting the total energy--momentum tensor, Eq.~(\ref{EQf3.06}),
into the continuity equation, Eq.~(\ref{EQf3.01a}), produces
\begin{equation}
\frac{\partial \rho_e}{\partial t}+ \nabla\cdot (n{\bf S}) = 0
\label{EQf3.07}
\end{equation}
for the $\alpha=0$ component.
The energy continuity equation, Eq.~(\ref{EQf3.07}), is manifestly false
because the two terms of Eq.~(\ref{EQf3.07}) depend on different powers
of the parameter $n$.
This is obvious if we compare Eq.~(\ref{EQf3.07}) with Poynting's
theorem, Eq.~(\ref{EQf2.06}), and it can also be proved by evaluating
the amplitudes of the components.
Then the conservation principle embodied in the continuity equation,
Eq.~(\ref{EQf3.01a}), is inconsistent with conservation of total energy
and conservation of total linear momentum, Eq.~(\ref{EQf3.01c}),
conservation of total angular momentum, Eq.~(\ref{EQf3.01b}),
and with the total energy--momentum tensor, Eq.~(\ref{EQf3.06}).
\par
\section{Discussion}
\par
The scientific literature contains a large number of theoretical and
experimental studies that claim to resolve the Abraham--Minkowski
controversy.
Here, we discuss how the results of the previous section
relate to examples from the prior work.
In particular, we find that the prior work stopped short of considering
the validity of the tensor total energy--momentum continuity equation.
\par
In an influential 1973 article, Gordon \cite{BIGord} uses a microscopic
model of the dielectric in terms of electric dipoles.
Assuming a dilute vapor in which the dipoles do not interact with each
other or their environment, Gordon averages the microscopic Lorentz
dipole force on an particle with linear polarizability $\alpha$
\begin{equation}
{\bf f}=\alpha \left ( ({\bf E}\cdot\nabla){\bf E}
+ \frac{d{\bf E}}{dt}\times{\bf B}
\right )
\label{EQf4.01}
\end{equation}
to obtain a macroscopic Lorentz force density that is integrated to
produce the material momentum density
\begin{equation}
{\bf g}_{matter}=(n-1)\frac{{\bf E}\times{\bf B}}{c} \, .
\label{EQf4.02}
\end{equation}
Gordon added the material momentum density to the Abraham momentum
density for the electromagnetic field to obtain
${\bf g}_T=(n/c){\bf E}\times{\bf B}$ for the total momentum density.
The corresponding total momentum, Eq.~(\ref{EQf1.01}), has been 
proved to be conserved in the situations that are considered here.
This result, alone, is insufficient to construct the tensor total
energy--momentum continuity equation.
If we also require that the total angular momentum and the total energy
are conserved, then we can construct the total energy--momentum
tensor, Eq.~(\ref{EQf3.06}).
It was shown in the preceding section that the resulting total energy
continuity equation, Eq.~(\ref{EQf3.07}), is false.
\par
In similarly influential work, Barnett \cite{BIAMC2} and
Barnett and Loudon \cite{BIBarnL} prove that the Abraham momentum is
the kinetic momentum and the Minkowski momentum is the canonical
momentum.
They assert that both results are correct because 1) they were derived
from fundamental principles and 2) the total momentum can be constructed
by adding a material momentum, different in each case, to the Abraham
kinetic momentum and to the Minkowski canonical momentum.
The sum of the kinetic momentum and its material momentum is equal to
the sum of the canonical momentum and its material momentum.
In each case, the sum is required to be the total linear momentum, which
is conserved.
Then, Eq.~(\ref{EQf1.01}) is the total linear momentum and we can
construct the total energy--momentum tensor, Eq.~(\ref{EQf3.06}), as
before.
Applying the four-divergence operator to the total energy--momentum
tensor as shown in Eq.~(\ref{EQf3.01a}), the component of the
energy--momentum continuity equation that relates to continuity of
energy, Eq.~(\ref{EQf3.07}), remains false.
\par
In a microscopic approach, the material momentum is modeled as the
kinematic momentum of individual particles of
matter \cite{BIPfei,BIObuk}.
The total energy--momentum tensor is assumed to be the sum of an
electromagnetic tensor and the dust tensor.
However, the dust energy--momentum tensor is usually applied to
a thermodynamically closed system consisting of non-interacting
neutral particles in an incoherent unimpeded flow.
Then we must derive the relationships between the electromagnetic and 
material subsystems from the conservation principles,
Eqs.~(\ref{EQf3.01b}) and (\ref{EQf3.01c}),
for the composite total energy--momentum tensor, Eq.~(\ref{EQf3.03}).
The resulting total energy--momentum tensor is unique and this tensor,
Eq.~(\ref{EQf3.06}), leads to provably false conservation principles,
as before.
\par
A number of experiments have been performed in order to determine the
momentum of the field in a dielectric, notably
Refs.~\cite{BIExp1,BIExp2,BIExp3}.
The results of electromagnetic momentum experiments, either the Minkowski
momentum or the Abraham momentum, are not sufficient to construct the
total energy--momentum tensor or the tensor total energy momentum continuity
equation.
We must rely on conservation principles to construct the total
energy--momentum tensor, Eq.~(\ref{EQf3.06}), which is unique.
Substituting the total enegy--momentum tensor into Eq.~(\ref{EQf3.01a})
results in a false statement of energy conservation, EQ.~(\ref{EQf3.07}).
\par
\section{Conclusions}
\par
The Abraham--Minkowski controversy cannot be resolved using the existing
theory because the principles of conservation are inconsistent in a
linear dielectric medium.
Specifically, the first row and first column of the total
energy--momentum tensor continuity equations can be constructed by
conservation of total energy, conservation of total linear momentum, and
conservation of total angular momentum.
However, the four-divergence of the total energy--momentum tensor 
produces a total energy continuity equation that is manifestly false.
This result shows that more than the assumption of a material subsystem
to add to an electromagnetic subsystem will be required to resolve
the Abraham--Minkowski controversy.
In future publications, we will show how to repair the conservation 
principles and related laws of physics in a dielectric medium.
\par

\end{document}